\documentclass[superscriptaddress,preprintnumbers,amsmath,amssymb,prd,nofootinbib,preprint]{revtex4-1}
\pdfoutput=1
\usepackage{graphicx}
\usepackage{epstopdf}
\usepackage{dcolumn}% Align table columns on decimal point
\usepackage{bm}% bold math
\usepackage{hyperref}
\usepackage{color}
\usepackage{amsmath}
\usepackage{cancel}
\usepackage{xpatch}

\begin{document}

%%%%%%%%%%%%%%%%%%%%%%%%%%%%%%%%%%%%%%%%%%%

\def\a{\alpha}
\def\b{\beta}
\def\c{\varepsilon}
\def\d{\delta}
\def\e{\epsilon}
\def\f{\phi}
\def\g{\gamma}
\def\h{\theta}
\def\k{\kappa}
\def\l{\lambda}
\def\m{\mu}
\def\n{\nu}
\def\p{\psi}
\def\q{\partial}
\def\r{\rho}
\def\s{\sigma}
\def\t{\tau}
\def\u{\upsilon}
\def\v{\varphi}
\def\w{\omega}
\def\x{\xi}
\def\y{\eta}
\def\z{\zeta}
\def\D{\Delta}
\def\G{\Gamma}
\def\H{\Theta}
\def\L{\Lambda}
\def\F{\Phi}
\def\P{\Psi}
\def\S{\Sigma}

\def\o{\over}
\def\beq{\begin{align}}
\def\eeq{\end{align}}
\newcommand{\gsim}{ \mathop{}_{\textstyle \sim}^{\textstyle >} }
\newcommand{\lsim}{ \mathop{}_{\textstyle \sim}^{\textstyle <} }
\newcommand{\vev}[1]{ \left\langle {#1} \right\rangle }
\newcommand{\bra}[1]{ \langle {#1} | }
\newcommand{\ket}[1]{ | {#1} \rangle }
\newcommand{\EV}{ {\rm eV} }
\newcommand{\KEV}{ {\rm keV} }
\newcommand{\MEV}{ {\rm MeV} }
\newcommand{\GEV}{ {\rm GeV} }
\newcommand{\TEV}{ {\rm TeV} }
\newcommand{\1}{\mbox{1}\hspace{-0.25em}\mbox{l}}
\newcommand{\headline}[1]{\noindent{\bf #1}}
\def\diag{\mathop{\rm diag}\nolimits}
\def\Spin{\mathop{\rm Spin}}
\def\SO{\mathop{\rm SO}}
\def\O{\mathop{\rm O}}
\def\SU{\mathop{\rm SU}}
\def\U{\mathop{\rm U}}
\def\Sp{\mathop{\rm Sp}}
\def\SL{\mathop{\rm SL}}
\def\tr{\mathop{\rm tr}}
\def\mpl{M_{\rm Pl}}

\def\IJMP{Int.~J.~Mod.~Phys. }
\def\MPL{Mod.~Phys.~Lett. }
\def\NP{Nucl.~Phys. }
\def\PL{Phys.~Lett. }
\def\PR{Phys.~Rev. }
\def\PRL{Phys.~Rev.~Lett. }
\def\PTP{Prog.~Theor.~Phys. }
\def\ZP{Z.~Phys. }

\def\dd{\mathrm{d}}
\def\ff{\mathrm{f}}
\def\BH{{\rm BH}}
\def\inf{{\rm inf}}
\def\ev{{\rm evap}}
\def\eq{{\rm eq}}
\def\SM{{\rm sm}}
\def\Mpl{M_{\rm Pl}}
\def\GeV{{\rm GeV}}
\def\Myr{\rm Myr}
\newcommand{\Red}[1]{\textcolor{red}{#1}}
\newcommand{\kh}[1]{\textcolor{blue}{\bf KH: #1}}
\newcommand{\DD}[1]{\textcolor{blue}{\bf DD: #1}}
\newcommand{\updm}{{\Delta'}}
\newcommand{\upex}{{h'}}

\title{
Higgs Parity, Strong CP %\\
and Dark Matter %\\
}
\author{David Dunsky}
\affiliation{Department of Physics, University of California, Berkeley, California 94720, USA}
\affiliation{Theoretical Physics Group, Lawrence Berkeley National Laboratory, Berkeley, California 94720, USA}
\author{Lawrence J. Hall}
\affiliation{Department of Physics, University of California, Berkeley, California 94720, USA}
\affiliation{Theoretical Physics Group, Lawrence Berkeley National Laboratory, Berkeley, California 94720, USA}
\author{Keisuke Harigaya}
\affiliation{School of Natural Sciences, Institute for Advanced Study, Princeton, New Jersey, 08540}
\affiliation{Department of Physics, University of California, Berkeley, California 94720, USA}
\affiliation{Theoretical Physics Group, Lawrence Berkeley National Laboratory, Berkeley, California 94720, USA}

\begin{abstract}
An exact spacetime parity replicates the $SU(2) \times U(1)$ electroweak interaction, the Higgs boson $H$, and the matter of the Standard Model. 
This ``Higgs Parity" and the mirror electroweak symmetry are spontaneously broken at scale $v' = \vev{H'} \gg \vev{H}$, yielding the Standard Model below $v'$  
with a quartic coupling that essentially vanishes at $v'$: $\lambda_{SM}(v') \sim 10^{-3}$.   The strong CP problem is solved as Higgs parity forces the masses of mirror quarks and ordinary quarks to have opposite phases.     
Dark matter is composed of mirror electrons, $e'$, stabilized by unbroken mirror electromagnetism. These interact with Standard Model particles via kinetic mixing between the photon and the mirror photon, which arises at four-loop level and is a firm prediction of the theory. 
Physics below $v'$, including the mass and interaction of $e'$ dark matter, is described by {\it one fewer parameter} than in the Standard Model. The allowed range of $m_{e'}$ is determined by uncertainties in $(\alpha_s, m_t, m_h)$, so that future precision measurements of these will be correlated with the direct detection rate of $e'$ dark matter, which, together with the neutron electric dipole moment, will probe the entire parameter space.
\end{abstract}

\date{\today}

\maketitle

\tableofcontents

\newpage
%%%%%%%%%%%%%%%%%%%
\section{Introduction}
%%%%%%%%%%%%%%%%%%%

For decades, a natural weak scale has been the key guide to constructing theories of physics beyond the Standard Model (SM), leading to new physics at or below the TeV scale.  However, so far LHC data points to an alternative view where the SM, with a highly perturbative Higgs boson, is the effective theory to extremely high energies.  In this case, the Higgs quartic coupling, $\lambda_{SM}$, exhibits extraordinary behavior taking an absolute value of $10^{-2}$ or less at energies above about $10^9$ GeV.  Indeed,  at  $2 \sigma$ 
\begin{align}
\label{eq:muc}
\lambda_{SM}(\mu_c) = 0,  
\end{align}
where $\mu_c \simeq (10^9 - 3 \times 10^{12})$ GeV~\cite{Buttazzo:2013uya} (see~\cite{Lindner:1988ww,Sher:1993mf,Altarelli:1994rb,Casas:1994qy,Espinosa:1995se,Casas:1996aq,Hambye:1996wb,Isidori:2001bm,Degrassi:2012ry} for earlier works).

In a recent paper \cite{Hall:2018let}, two of us introduced a new framework, ``Higgs Parity", to understand this behavior of the SM quartic.  A $Z_2$ symmetry replicates the $SU(2)$ gauge group of the SM, $SU(2) \leftrightarrow SU(2)'$, with the Higgs sector transforming as $H(2,1) + H'(1,2)$ under $(SU(2), SU(2)')$ and is spontaneously broken at scale $v' = \vev{H'} \gg \vev{H}$.  A SM Higgs sector much lighter than $v'$ requires a fine-tuning that makes the Higgs a pseudo Nambu-Goldstone boson of an accidental $SU(4)$ symmetry. The SM Higgs quartic coupling then arises only at the loop level, so that 
\begin{align}
\label{eq:v'}
|\lambda_{SM}(v')| = {\cal O} (10^{-3})
\end{align}
and $\mu_c$ is close to $v'$.

The strong CP problem~\cite{tHooft:1976rip} can be addressed by introducing spacetime parity~\cite{Beg:1978mt,Mohapatra:1978fy}, and a viable theory was first constructed by Babu and Mohapatra~\cite{Babu:1989rb}.
Higgs Parity provides a solution to the strong CP problem if it is promoted to a spacetime parity, $P$, and {\it does not} replicate QCD \cite{Hall:2018let}.  

Thus, simple theories with Higgs Parity can simultaneously solve the strong CP problem and account for the extraordinary behavior of the SM quartic, making them a significant competitor to axion theories~\cite{Peccei:1977hh,Peccei:1977ur}.  However, without a Weakly Interacting Massive Particle or an axion, the nature of dark matter (DM) in these theories becomes pressing.  In this paper we show that such theories have a built-in DM candidate provided $P$ replicates the entire electroweak gauge group as well as the quarks and leptons.  DM is composed of mirror electrons and positrons, stabilized by an unbroken $U(1)_{QED}'$. The mirror baryon made of three mirror up quarks is also stable. However, a strong upper bound on the abundance of exotic hadrons, made both of mirror up quarks and SM quarks, requires that only a very small fraction of DM can arise from such mirror baryons. The suppression of the mirror up quark abundance requires that $e'$ is produced non-thermally as we will discuss. 

In this paper we study a theory that has the same number of parameters as the SM.  Remarkably, these parameters allow us to compute the DM mass, its self interactions, and its interactions with SM particles. 

At energies above $\mu_c$, the gauge group is $SU(3) \times (SU(2) \times U(1)) \times  (SU(2)' \times U(1)')$ and parity ensures three independent gauge couplings, as in the SM.  The Higgs potential involves three parameters, rather than the two of the SM; two describe the two symmetry breaking scales of $\langle H' \rangle = v'$ for $SU(2)' \times U(1)' \rightarrow U(1)_{QED}'$ and $\langle H \rangle = v$ for $SU(2) \times U(1) \rightarrow U(1)_{QED}$, while the third is irrelevant to us since it fixes the mirror Higgs mass, $m_{h'}$.  The Yukawa coupling matrices of the mirror sector are the complex conjugate of those of the SM sector.  Thus mirror quark and charged lepton masses are larger than those of the SM by $v'/v$ (and calculable renormalization factors) and the strong CP parameter $\bar{\theta} = 0$.  Since the gauge and Yukawa couplings in our theory are the same as in the SM, the change in parameter space may be described by
\begin{align}
\label{eq:params}
\{ m_h, v, \bar{\theta} \} \rightarrow  \{ v, v', m_{h'} \} \rightarrow  \{ v, v'\}. 
\end{align}
The last stage signifies that the mirror Higgs mass has no effect on any experimental observable.  Particle physics and dark matter physics are described by one parameter less than in the SM; however, additional physics is required to understand the DM abundance.
There could be extra parameters in the mirror neutrino masses, but $\nu'$ are very heavy and play no role in this paper.

The mass and interaction strength of DM particles are not free parameters.
The mirror electron $e'$ interacts with SM particles via $U(1)$ kinetic mixing, which arises at four-loop level and is a prediction of the theory.
The mirror electroweak scale $v'$, and hence the mass of DM $m_e(v'/v)$,  is fixed once the SM Higgs mass, the top quark mass and the strong coupling constant are measured with a sufficient accuracy. The theory thus predicts a tight correlation between these three parameters and the direct detection rate of DM.

Although the strong CP parameter vanishes at the renormalizable level, a non-zero value arises from a dimension-6 interaction between the Higgs and gluon fields. Assuming a cut-off scale at or below the Planck mass, a neutron electric dipole moment is expected to be observed in near future experiments.

In section~\ref{sec:Z2} we review how a $Z_2$ symmetry of the Higgs sector, $H(2,1) + H'(1,2)$, spontaneously broken by $\langle H' \rangle = v'$, leads to $\lambda_{SM}(v') = 0$ at tree level. In section~\ref{sec:MirrorEW} we describe the Lagrangian of the theory and show that the strong CP problem is solved. We compute the four-loop correction to the $U(1)$ kinetic mixing and the relation between the SM parameters and $v'$.
In section~\ref{sec:ObsCon}, observational constraints on mirror DM is discussed, and the correlation between SM parameters and the direct detection rate of DM is shown. In section~\ref{sec:nonthermal}, non-thermal production of mirror electrons is discussed.

%%%%%%%%%%%%%%%%%%%%%%%%%%
\section{Vanishing Higgs Quartic from a $Z_2$ Symmetry}
\label{sec:Z2}
%%%%%%%%%%%%%%%%%%%%%%%%%%%
In this section we review the framework of \cite{Hall:2018let} that yields the near vanishing of the SM Higgs quartic coupling at a high energy scale. 
Consider a $Z_2$ symmetry that exchanges the $SU(2)$ weak gauge interaction with a new $SU(2)'$ gauge interaction, and the Higgs field $H(2,1)$ with its partner $H'(1,2)$, where the brackets show the $(SU(2), SU(2)')$ representation. The scalar potential for $H$ and $H'$ is given by
\begin{align}
\label{eq:potential}
V(H,H') = - m^2 (H^\dagger H + H'^\dagger H') + \frac{\lambda}{2} (H^\dagger H + H'^\dagger H')^2 + \lambda' H^\dagger H H'^\dagger H' .
\end{align}
We assume that the mass scale $m$ is much larger than the electroweak scale. With $m^2$ positive, the $Z_2$ symmetry is spontaneously broken and $H'$ acquires a large vacuum expectation value of $\vev{H'} = v'$, with $v'^2 = m^2/\lambda$. After integrating out $H'$ at tree-level, the Low Energy potential in the effective theory for $H$ is
\begin{align}
\label{eq:potentialLE}
V_{LE}(H) = \lambda' \; v'^2  \; H^\dagger H - \lambda' \left(1  + \frac{\lambda'}{2 \lambda} \right) (H^\dagger H)^2 .
\end{align}
To obtain the hierarchy $\vev{H} = v \ll v'$, it is necessary to tune $\lambda'$ to a very small value $\lambda'  \sim - v^2/v'^2$; the quartic coupling of the Higgs $H$, $\lambda_{\rm SM}$, is then extremely small.  

The vanishing quartic can be understood by an accidental $SU(4)$ symmetry under which $(H, H')$ is in a fundamental representation. For $|\lambda'| \ll 1$,  necessary for $v \ll v'$, the potential in Eq.~(\ref{eq:potential}) becomes $SU(4)$ symmetric.
 After $H'$ obtains a vacuum expectation value, the Standard Model Higgs is understood as a Nambu-Goldstone boson with a vanishing potential.
 Note that in this limit of extremely small $\lambda'$, the vacuum alignment in the SU(4) space is determined by the Coleman-Weinberg potential.  The top contribution beats the gauge contribution so that the true vacuum is the asymmetric one, where the entire condensate lies in $H'$ (or in $H$, which is physically equivalent). (The $SU(4)$ symmetry implies that the Higgs boson contribution to the Coleman-Weinberg potential does not affect the vacuum orientation.)
 
Below the scale $v'$, quantum corrections from SM particles renormalize the quartic coupling, and it becomes positive.
From the perspective of running from low to high energies, the scale at which the SM Higgs quartic coupling vanishes, $\mu_c$ of (\ref{eq:muc}), is identified with $v'$,   $v' \simeq \mu_c$. The threshold correction to $\lambda_{\rm SM}(v')$ is calculated in the next section.

Although the scale $v'$ is much smaller than the Planck scale and the typical unification scale, the theory is no more fine-tuned than the Standard Model because of the $Z_2$ symmetry.
The required fine-tuning is
\begin{align}
\frac{m^2}{\Lambda^2} \times \frac{v^2}{m^2} = \frac{v^2}{\Lambda^2},
\end{align}
where the first factor in the left hand side is the fine-tuning to obtain the scale $m$ much smaller than the cut off scale $\Lambda$, and the second one is the fine-tuning in $\lambda'$ to obtain the electroweak scale from $m$. The total tuning is the same as in the Standard Model, $v^2 / \Lambda^2$, and may be explained by environment requirements \cite{Agrawal:1997gf,Hall:2014dfa}.

%We assume that the $Z_2$ symmetry is exact above the scale $v'$, and is only broken when $H'$ condenses.
It is considered that a global symmetry is always explicitly broken in quantum gravity~\cite{giddings1988loss,coleman1988there,gilbert1989wormhole,Banks:2010zn,Harlow:2018jwu,Harlow:2018tng}.
%If, however, the $Z_2$ symmetry is a global symmetry, it may be broken near the Planck scale since quantum gravity conjectures forbid any global, discrete symmetries \cite{giddings1988loss,coleman1988there,gilbert1989wormhole,Banks:2010zn,Harlow:2018jwu,Harlow:2018tng}.
We may  gauge the $Z_2$ symmetry so that it remains exact above the scale $v'$~\cite{Dine:1992ya,Choi:1992xp}, and is only spontaneously broken when $H'$ condenses.

In \cite{Hall:2018let} it was shown that the strong CP problem~\cite{tHooft:1976rip} is solved if the $Z_2$ symmetry includes space-time parity and leaves the QCD interaction invariant.  In this paper we choose to have $Z_2$ replicate the full electroweak interaction, so that there is an unbroken mirror QED symmetry that stabilizes light mirror matter~\cite{Barr:1991qx} allowing it to be DM~\cite{Hall:talk}.

%%%%%%%%%%%%%%%%%%%%%%%%%%
\section{The Mirror Electroweak Theory}
\label{sec:MirrorEW}
%%%%%%%%%%%%%%%%%%%%%%%%%%%
In this paper we study a theory where the electroweak gauge group, $SU(2) \times U(1)$, is replicated by a parity symmetry, while the QCD interaction is invariant; thus the gauge group is $SU(3) \times (SU(2) \times U(1)) \times (SU(2)' \times U(1)')$.  The Standard Model matter ($q, \bar{u}, \bar{d}, \ell, \bar{e})$ and Higgs are neutral under $SU(2)' \times U(1)'$, and the action of parity is
\begin{align}
\label{eq:Paction}
\bar{x}  \; \; \leftrightarrow \;\; &  - \bar{x}  \nonumber \\  
 SU(2) \times U(1) \; \; \leftrightarrow \; \; & SU(2)' \times U(1)' \nonumber \\
 q, \bar{u}, \bar{d}, \ell, \bar{e} \; \; \leftrightarrow \; \; & (q', \bar{u}', \bar{d}', \ell', \bar{e}')^\dagger \nonumber \\
 H \; \; \leftrightarrow \; \; & H',
\end{align}
where matter is described by 2-component Weyl fields.

%%%%%%%%%
\subsection{Renormalizable interactions}
%%%%%%%%%

The most general gauge and parity invariant Lagrangian up to dimension 4 is given by 
\begin{align}
\label{eq:L4}
{\cal L}_4 \; =& \; {\cal L}_{KE} - \frac{\epsilon_B}{2} \; B^{\mu \nu} B'_{\mu \nu} + {\cal L}_Y - V(H,H') 
\end{align}
where ${\cal L}_{KE}$ contains canonical kinetic energies for all fields, $\epsilon_B$ describes kinetic mixing between ordinary and mirror hypercharge and the QCD $\theta$ parameter is absent due to parity.   $V(H,H')$ is the Higgs potential of (\ref{eq:potential}), and Yukawa couplings are described by
\begin{align}
\label{eq:Yuk}
{\cal L}_Y =&  (q \, y_u \bar{u}) H^\dag +  (q \, y_d \bar{d}) H + (\ell \, y_e \bar{e}) H  +  (q'  y^*_u \bar{u}') H^{'\dag} +(q'  y^*_d  \bar{d}') H'  + (\ell'  y^*_{e}\bar{e}')  H'   + {\rm h.c.}
\end{align}
where $y_{u,d,e}$ are the SM $3 \times 3$ Yukawa coupling matrices and parity implies that the mirror Yukawa matrices are the complex conjugate of the SM ones.  

As $V(H,H')$ has three parameters, this theory possesses a single extra parameter compared to the SM.  The analysis of the previous section applies: without loss of generality, in the limit of small $\lambda'$, the vacuum has $\vev{H}=v \ll \vev{H'} = v'$, and $\lambda_{SM}(v')=0$ at tree level.  In this theory the observed values of $G_F$ and the Higgs mass determine $v$ and $v'$, and the third parameter of the Higgs potential determines the mirror Higgs mass and is irrelevant for physics below the scale $v'$.

%%%%%%%%%
\subsection{Strong CP problem}
%%%%%%%%%
The $6 \times 6$ mass matrices for the $(u,d)$ quarks of the two sectors are
\begin{align}
\label{eq:6X6BC}
{\cal M}_{u,d} = \begin{pmatrix}
y^*_{u,d} \, v'& 0\\
0 & y_{u,d} \, v
\end{pmatrix}.
\end{align}
Mirror and standard quarks give equal and opposite phases to the determinant of their mass matrices, so that $\bar{\theta} = 0$ at tree level. Loop corrections give rise to $\bar{\theta} \sim O(10^{-16})$ as in the Standard Model~\cite{Ellis:1978hq}, corresponding to a neutron electric dipole moment of order $10^{-31}$ e cm, so that the strong CP problem is solved.  This method of using parity to solve the strong CP problem was invented by Barr, Chang and Senjanovic \cite{Barr:1991qx}. The vanishing Dirac mass limit of the model by Babu and Mohapatra~\cite{Babu:1989rb} reduces to this method.

The effective field theory contains the Higgs Parity even, dimension 6 operator
\begin{align}
\label{eq:HHGG}
{\cal L}_6 = \frac{C}{M_{Pl}^2} (|H^2| - |H'|^2) \, G \tilde{G},
\end{align}
where $G$ is the field strength of $SU(3)_c$, $M_{Pl} = 2.4 \times 10^{18}~{\rm GeV}$ is the reduced Planck mass, and $C$ is a dimensionless coupling. Condensation of $H'$ yields the strong CP phase
\begin{align}
\theta = 32\pi^2 C \left(\frac{v'}{M_{Pl}} \right)^2  = 5\times 10^{-11}  C \left( \frac{v'}{10^{12}~{\rm GeV}} \right)^2.
\end{align}
We will find that DM places a lower bound on $v'$, giving a result for $\theta$ close to the experimental constraint, $\theta < 10^{-10}$~\cite{Crewther:1979pi,Baker:2006ts,Graner:2016ses}, that could be discovered in on-going searches for the neutron electric dipole moment \cite{Lamoreaux:2009zz, Baker:2016bzc, Tsentalovich:2014mfa}.

The strong CP problem can be also solved by a CP symmetry, which forbids the theta term.
Since CP symmetry also requires Yukawa couplings to be real, the CKM phase is obtained by spontaneous breaking of CP.  A one-loop quantum correction to the strong CP phase can be suppressed by sophisticated setups~\cite{Nelson:1983zb,Barr:1984qx,Bento:1991ez,Hiller:2001qg}. In the parity solution, parity does not require Yukawa couplings to be real and the CKM matrix is easily reproduced. 

%%%%%%%%%
\subsection{Kinetic Mixing at 4 loops}
\begin{figure}[tb]
\centering
\includegraphics[width=0.7\textwidth]{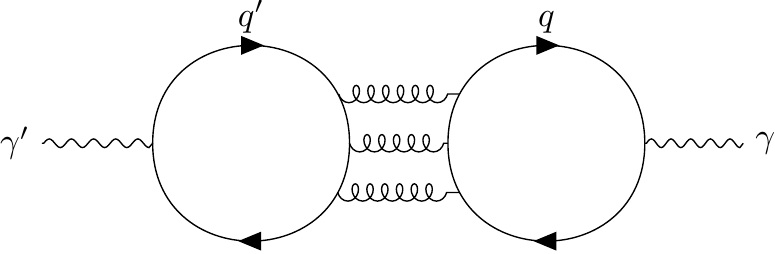} %
\caption{Four-loop diagram that gives rise to kinetic mixing between sectors.}
\label{fig:epsilon_diagram}
\end{figure}

\begin{figure}[tb]
\centering
\includegraphics[width=0.7\textwidth]{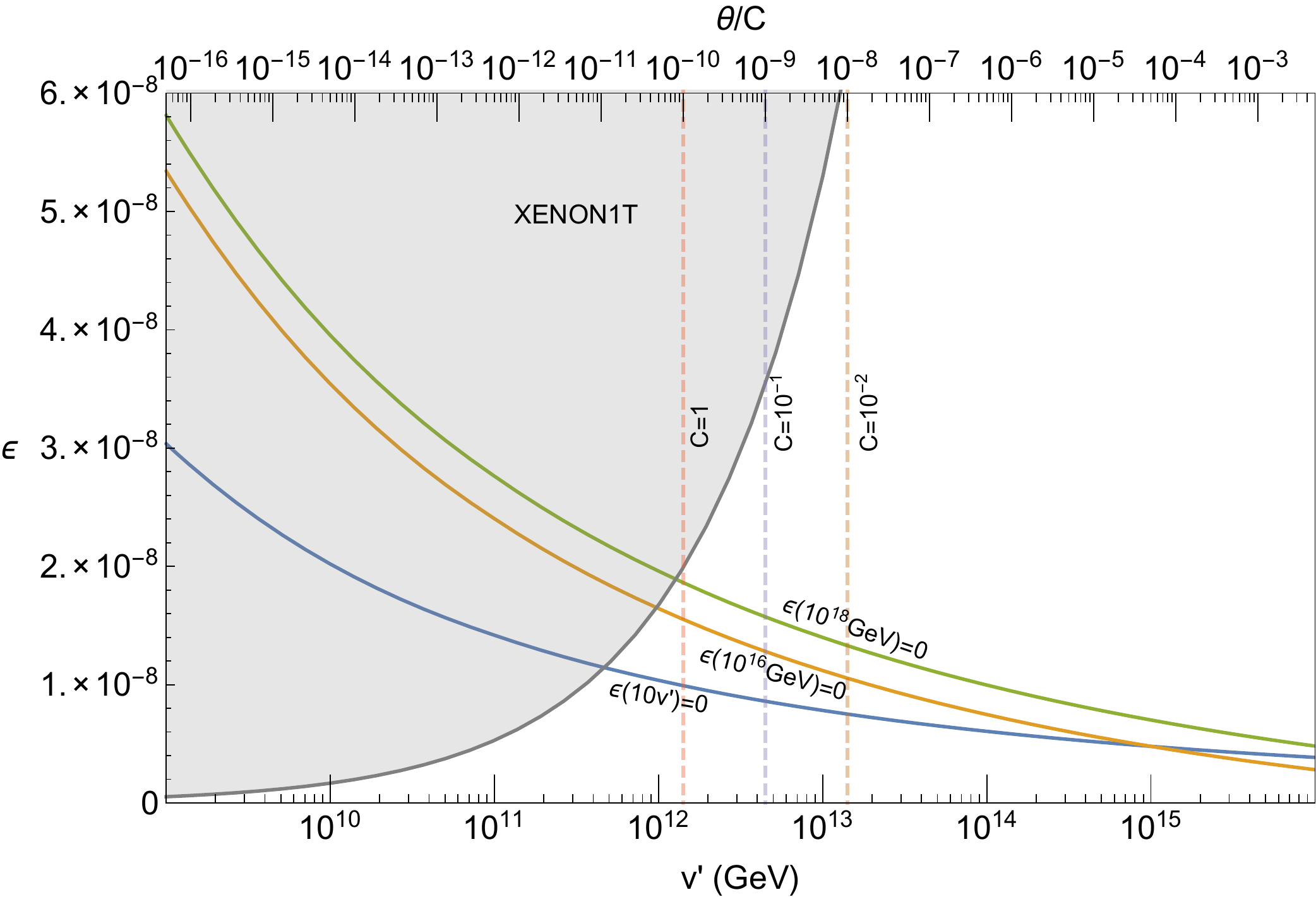} %
\caption{The prediction for the kinetic mixing parameter $\epsilon$ as a function of $v'$, for three values of the cutoff of the theory.  If DM is $e'$, the shaded region is excluded by the direct detection limit of XENON1T. For various values of the coupling $C$, defined in (\ref{eq:HHGG}), the present limit on the neutron electric dipole moment excludes the region to the right of the vertical lines. }
\label{fig:epsilon}
\end{figure}

%%%%%%%%%
Kinetic mixing between the standard and mirror sectors is induced at four loops by the shared color charge of standard and mirror quarks, as shown in Fig. \ref{fig:epsilon_diagram}.
We may directly compute the kinetic mixing between the SM photon and the mirror photon by projecting the external gauge field into the massless combination. The renormalization group equation of the kinetic mixing parameter can be read off from the four-loop beta function of QCD~\cite{vanRitbergen:1997va},
\begin{align}
\frac{{\rm d}}{{\rm dln}\mu} \left( \frac{\epsilon}{e^2} \right)= \frac{g_3^6}{(4\pi)^8} \left(  - \frac{1760}{27} + \frac{1280}{9}\zeta(3) \right) \sum_{i j}q_i q'_j.
\label{eq:4loop}
\end{align}
Here $i$ runs over all the quark charges, $q_i$, while $j$ is summed only over mirror quarks with mass below the scale $\mu$.
The prediction for $\epsilon$ is shown in Fig.~\ref{fig:epsilon} as a function of $v'$. Here we take the boundary condition $\epsilon(\Lambda)=0$, where $\Lambda$ is the UV cutoff of our theory.  This results if either $U(1)$ is incorporated in a non-Abelian factor above $\Lambda$ providing any particles carrying both $U(1)$ charges are much heavier than $\Lambda$. The three curves correspond to  $\Lambda = 10v', 10^{16} \; \GeV$ and $10^{18}$ GeV.  Even with $\Lambda$ as low as $v'$ there are large logarithms, such as $
\ln v'/m_{u'}$, so that the the solution of (\ref{eq:4loop}) is expected to dominate over finite contributions. The result, $\epsilon = O(10^{-8})$, is important for placing a limit on the mass of $e'$ from DM direct detection, and the large numerical factor of (\ref{eq:4loop}) plays a crucial role.
%%

%%%%%%%%%
\subsection{Neutrino Masses}
%%%%%%%%%
Standard and mirror neutrinos obtain mass from operators of dimension 5 
\begin{align}
\label{eq:high_dimnu}
{\cal L}_5 = (\ell \, \eta \, \ell) \frac{H^2}{M_M}   +  (\ell' \, \eta^* \, \ell') \frac{H'^{\dagger2}}{M_M}  + (\ell \, \xi \, \ell') \frac{H H'^\dagger}{M_D} 
+ {\rm h.c.}
\end{align}
where $M_{M,D}$ are large mass scales and $\eta$ and $\xi$ are $3 \times 3$ dimensionless flavor matrices.
Taking $M_D \gg M_M$, where the mixing between $\nu'$ and $\nu$ is small, so that $m_{\nu'}/m_\nu \simeq (v'/v)^2$, gives
\begin{align}
\label{eq:mnu}
m_{\nu'} \simeq 10^{11} \; \GeV \left( \frac{m_\nu}{0.03 \; \mbox{eV}} \right) \left( \frac{v'}{10^{13}~{\rm GeV}} \right)^2.
\end{align}
%%

%%%%%%%%%
\subsection{Threshold correction to $\lambda(v')$}
%%%%%%%%%

We start from the one-loop Coleman-Weinberg potential of the theory above the mirror electroweak scale,
\begin{align}
V_{\rm tree} = & \lambda \left( |H|^2 + |H'|^2 \right)^2 + \lambda'|H|^2 |H'|^2 - m^2 (|H|^2 + |H'|^2),\\
V_{1-{\rm loop}} = & c |H|^4 \, {\rm ln} \frac{|H|}{M} + c |H'|^4 \, {\rm ln} \frac{|H'|}{M},\nonumber \\
c \equiv & - \frac{3}{8\pi^2}y_t^4 + \frac{3}{128\pi^2}(g^2 + {g'}^2)^2 + \frac{3}{64\pi^2}g^4,
\end{align}
where $M$ is an arbitrary scale. A change of $M$ can be absorbed by a change of $\lambda$. We take $M$ to be the vev of $H'$, which is given by
\begin{align}
v' \equiv \vev{H'} = \sqrt{ \frac{2 m^2}{4 \lambda + c}}.
\end{align}
After integrating out $H'$, the potential of $H$, to the leading order in $c$ and $\lambda'$, is given by
\begin{align}
V(H) \simeq {v'}^2(\lambda' - \frac{c}{2}) \, |H|^2 + (\frac{3}{4} c - \lambda' + c~{\rm ln} \frac{|H|}{v'})\, |H|^4.
\end{align}
To obtain the electroweak scale much smaller than $v'$, $\lambda' \simeq c /2$ is required. Then the Higgs potential is given by
\begin{align}
V(H) / |H|^4  \simeq \frac{c}{4} \, ( 1 + 4 \, {\rm ln} \frac{|H|}{v'} ).
\end{align}

We match this potential to the one-loop Coleman-Weinberg potential of the SM,
\begin{align}
V_{\rm SM}(H)/ |H|^4  = &\lambda_{\rm SM}(\mu)
- \frac{3}{16\pi^2} y_t^4\left( {\rm ln}\frac{y_t^2 |H|^2}{ \mu^2} - \frac{3}{2}  \right) \\
&+ \frac{3}{256\pi^2} (g^2 + {g'}^2)^2 \left( {\rm ln}\frac{ (g^2 + {g'}^2) |H|^2 / 2}{ \mu^2} - \frac{3}{2}  \right)
+ \frac{3}{128\pi^2} g^4 \left( {\rm ln}\frac{ g^2 |H|^2 / 2}{ \mu^2} - \frac{3}{2}  \right), \nonumber
\end{align}
where we take the $\overline{\rm MS}$ scheme. 
By matching $V_{\rm SM}(H)$ to  $V(H)$ with $\mu = v'$, we obtain
\begin{align}
\lambda_{\rm SM}(v') \simeq - \frac{3}{8\pi^2} y_t^4 \, {\rm ln} \frac{e}{y_t} + \frac{3}{128\pi^2} (g^2 + {g'}^2)^2 \, {\rm ln} \frac{e}{\sqrt{(g^2 + {g'}^2) / 2 }} + \frac{3}{64\pi^2} g^4 \, {\rm ln} \frac{e}{ g/ \sqrt{2}}.
\end{align}
A numerical evaluation shows that $\lambda_{\rm SM}(v')$ is negative and $O(10^{-3})$.

In Table~\ref{tab:vp}, we show the prediction for $v'$ for a wide variety of $(m_t,\alpha_s(m_Z))$.
To compute the running of the quartic coupling we follow the computation in~\cite{Buttazzo:2013uya}, adding the contribution from the mirror quarks to the running of the $SU(3)_c$ coupling constant at one-loop level.%
\footnote{We estimate the uncertainty due to the one-loop approximation by shifting the mirror quark thresholds by an $O(1)$ factor, and find that the uncertainty on the prediction of $v'$ is less than 10\%.}
For each $(m_t,\alpha_s(m_Z))$, the range of the prediction corresponds to the 1-sigma uncertainty in the measured Higgs mass, $m_h = (125.18\pm 0.16$) GeV. The $\overline{\rm MS}$ quartic coupling at $\mu=m_t$ reported in~\cite{Buttazzo:2013uya} has a theoretical uncertainty of $0.0003$, equivalent to a shift of the Higgs mass by 0.15 GeV, which is comparable to the uncertainty in the measurement of the Higgs mass.
The reference values of $(m_t,\alpha_s(m_Z))$ corresponds to the central values and the $1\mathchar`-2 \sigma$ ranges, derived from the experimental results $m_t= (173.0 \pm0.4)$ GeV, $\alpha_s(m_Z)=0.1181\pm0.0011$~\cite{Tanabashi:2018oca}.

\begin{table}[ht]
\caption{The prediction for $v'$ for $m_h=(125.18\pm 0.16)$ GeV.}
\begin{center}
\begin{tabular}{|c|c|c|c|c|c|}
\hline
$\alpha_s(m_Z) \backslash m_t$ & 173.8 GeV & 173.4 GeV & 173.0 GeV & 172.6 GeV & 172.2 GeV \\ \hline

0.1159 & $(2.6\mathchar`-3.4)\times 10^{9}$ & $(4.9\mathchar`-6.9)\times 10^{9}$ & $(1.0\mathchar`-1.5)\times 10^{10}$ & $(2.5\mathchar`-3.8) \times 10^{10}$ & $(0.67\mathchar`-1.1)\times 10^{11}$ \\

0.1170 & $(4.8\mathchar`-6.7)\times 10^{9}$ & $(1.0\mathchar`-1.5)\times 10^{10}$ & $(2.4\mathchar`-3.7)\times 10^{10}$ & $(0.66\mathchar`-1.1) \times 10^{11}$ & $(2.2\mathchar`-4.0)\times 10^{11}$ \\
 
0.1181 & $(1.0\mathchar`-1.5)\times 10^{10}$ & $(2.4\mathchar`-3.7)\times 10^{10}$ & $(0.65\mathchar`-1.1)\times 10^{11}$ & $(2.2\mathchar`-4.0) \times 10^{11}$ & $(0.95\mathchar`-2.1)\times 10^{12}$ \\
 
0.1192 & $(2.3\mathchar`-3.6)\times 10^{10}$ & $(0.64\mathchar`-1.1)\times 10^{11}$ & $(2.1\mathchar`-4.0)\times 10^{11}$ & $(0.96\mathchar`-2.1) \times 10^{12}$ & $(0.66\mathchar`-1.9)\times 10^{13}$ \\

0.1203 & $(0.63\mathchar`-1.0)\times 10^{11}$ & $(2.1\mathchar`-4.0)\times 10^{11}$ & $(0.97\mathchar`-2.2)\times 10^{12}$ & $(0.70\mathchar`-2.1) \times 10^{13}$ & $(1.2\mathchar`-7.3)\times 10^{14}$ \\ \hline

\end{tabular}
\end{center}
\label{tab:vp}
\end{table}%

\begin{figure}[tb]
\centering
\includegraphics[width=0.7\textwidth]{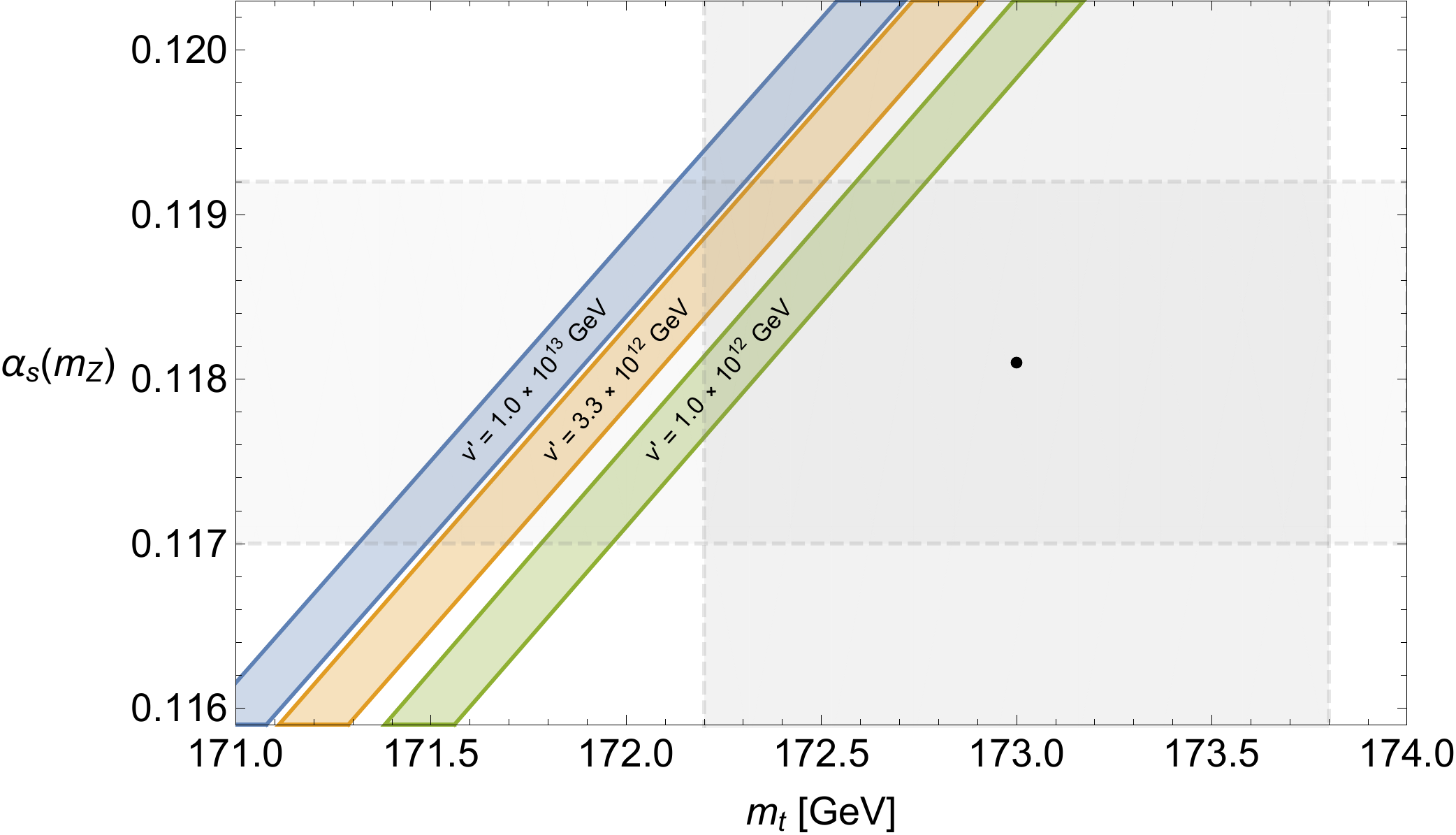} %
\caption{The prediction for $v'$ as a function of $m_t$ and $\alpha_s(m_Z)$. The thickness of the $v'$ contours is due to the uncertainty in the Higgs boson mass. The gray shaded rectangles show the current experimental values for $m_t$ at $2\sigma$ and $\alpha_s(m_Z)$ at $1 \sigma$. More precise measurements of these quantities will hone in on $v'$.}
\label{fig:alphaSvsMtopPlot}
\end{figure}
%%

%%%%%%%%%%%%%%%%%%%%%%%%%%%%%%%%%%%%
\section{Observational Constraints on $e'$ and $u'$ Dark Matter}
\label{sec:ObsCon}
%%%%%%%%%%%%%%%%%%%%%%%%%%%%%%%%%%%%%

\begin{figure}[tb]
\centering
\includegraphics[width=0.7\textwidth]{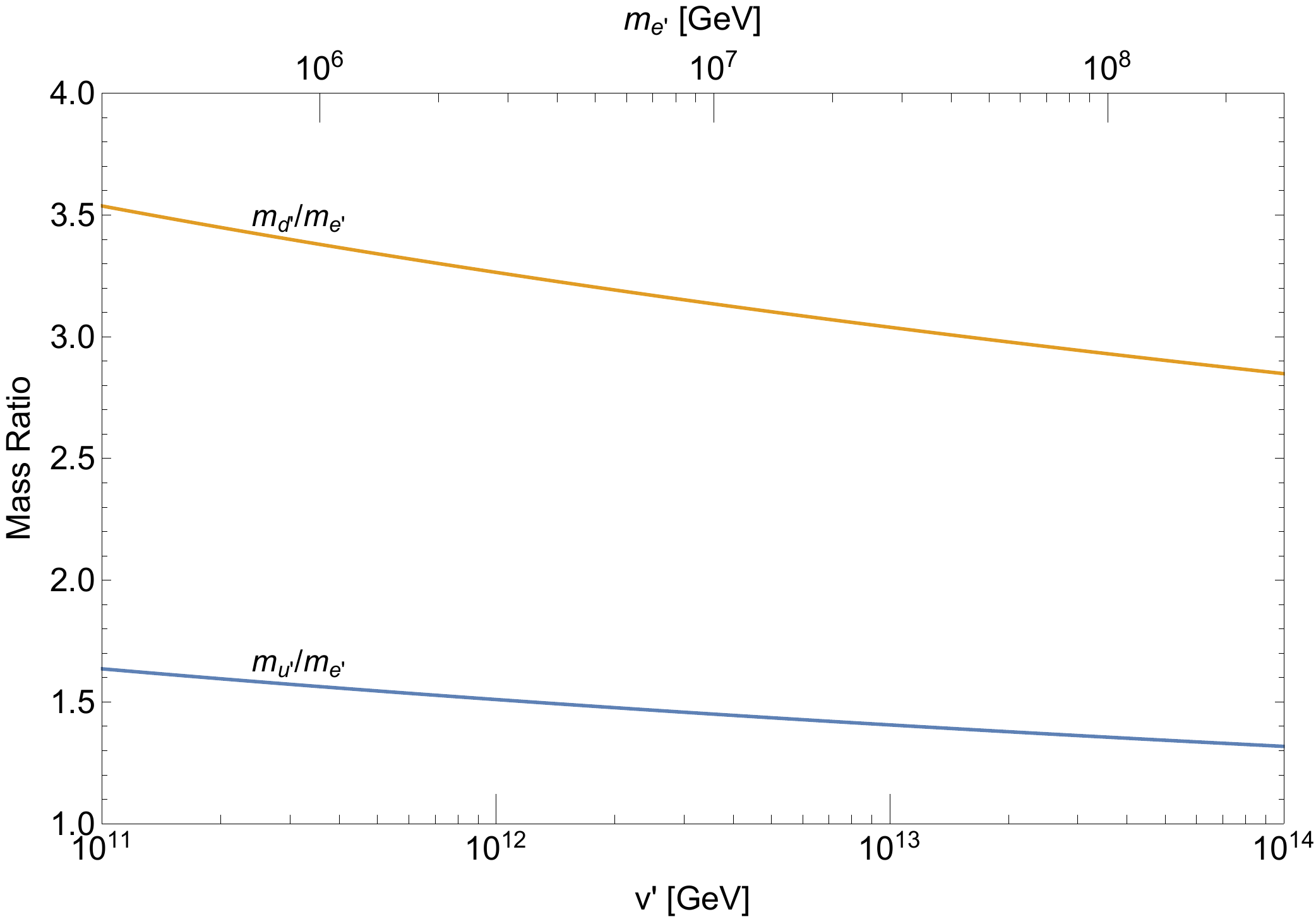} %
\caption{Masses of the lightest charged mirror fermions, $e', u'$ and $d'$.}
\label{fig:mirrorSpectrum}
\end{figure}

The mirror fermions acquire a mass $m_{f'} = y_{f'}v'$ from the vacuum expectation value of the mirror Higgs, $v'$. The $Z_2$ symmetry sets $y_{f'} = y_f$ at the scale $\mu = v'$, so that mirror fermion masses are larger than their SM counterparts by a factor of approximately $v'/v$, as shown in Fig. \ref{fig:mirrorSpectrum}. Note that the Yukawa couplings of mirror quarks  run faster than those of mirror leptons due to their additional $SU(3)$ charge. Consequently, the mirror electron and mirror up quark masses are nearly degenerate at large $v'$.

%%

%%%%%%%%%%%%%%%%%%
\subsection{Hadronization of $u'$}
\label{subsec:u'DM}
%%%%%%%%%%%%%%%%%%

After the QCD phase transition, $u'$ quarks form bound states by combining with other colored particles, namely, they hadronize.   Hadronization of massive colored particles and their subsequent evolution was investigated in~\cite{Kang:2006yd}.
%It was pointed out in~\cite{DeLuca:2018mzn} that the thermal relic of hadrons composed of a new colored particle may be dark matter.

Since the ordinary SM quarks, which we collectively denote as $q$, are much more abundant than $u'$, the $u'$ first form bound states $u'qq$ and $u'\bar{q}$, having $B'$ number of 1/3. These hadrons have a large radius $\sim \Lambda_{\rm QCD}^{-1}$ and, if sufficiently dense, can efficiently scatter with each other to rearrange constituents~\cite{Kang:2006yd}. In particular, states with $B'=2/3$ are formed by
\begin{align}
\label{eq:rearrangement1}
u'qq + u'qq \rightarrow (u'u' q)^* + qqq,~ \;
u'qq + u'\bar{q} \rightarrow (u'u'q)^*,~ \;
u'\bar{q} + u'\bar{q} \rightarrow (u'u'q)^* + \bar{q}\bar{q}\bar{q}
\end{align}
and similarly there is a processing of $B'=-1/3$ hadrons to those with $B'=-2/3$ by the corresponding antiparticle reactions.
The rearrangements may involve emission of pions, which we omit here and hereafter.
In addition, rearrangements can form $B'=0$ mesons containing $u'\bar{u}'$
\begin{align}
\label{eq:rearrangement2}
u'qq + \bar{u}\bar{q}\bar{q} \rightarrow (u'\bar{u}')^*,~
u'qq + \bar{u}'q \rightarrow (u'\bar{u}')^* + qqq,~
u'\bar{q} + \bar{u}' q \rightarrow  (u'\bar{u}')^*.
\end{align}
The two $u'/\bar{u}'$ in the $B'=\pm 2/3,0$ hadrons are initially at a distance of $O(\Lambda_{\rm QCD}^{-1})$ and in excited states denoted by a superscript $*$. They lose energy by emitting hadrons and fall into the ground state where the two $u'/\bar{u}'$ are bound by a Coulomb potential and have a separation of $O\left(\left(m_u' \alpha_3\right)^{-1} \right)$.
Once they fall into the ground state, mesons composed of $u'$ and $\bar{u}'$ decay via annihilation into SM hadrons, depleting the $u'$ number.

Once $B'=\pm2/3$ baryons form, further rearrangement reactions lead to the production of baryons with $B'=\pm1$
\begin{align}
\label{eq:rearrangement3}
u'u' q + u' \bar{q} \rightarrow (u'u'u')^*,~ \hspace{0.3in}
u'u' q + u'qq \rightarrow (u'u'u')^* + qqq
\end{align}
and similarly for the production of antibaryons via the antiparticle reactions.
Processes such as $u'u'q + \bar{u}'q \rightarrow u'\bar{u}'+ u'qq$ do not occur as they require the separation of deeply bound $u'$s in the first baryon. The excited states $(u'u'u')^*$ fall into the ground state $u'u'u'$, which has a radius of $O\left(\left(m_u' \alpha_3\right)^{-1} \right)$. Because of the small radius, the $u'u'u'$ do not participate in further rearrangements, and the $u'$ number is frozen once it forms the $u'u'u'$ state~\cite{DeLuca:2018mzn}.%
\footnote{This should be compared with the result of~\cite{Harigaya:2016nlg}. There it is assumed that the mass of the constituent is small enough so that the ground state is easily excited to a state with a large radius. As a result the depletion of the $u'$ number is not prevented by the formation of the ground state, and the DM abundance is much smaller than the abundance of the constituent before the phase transition, allowing a DM mass above the unitarity limit~\cite{Griest:1989wd}.}

In summary, the initial $u'$ have three possible fates.  They can:
1) Form hadrons including one or two $u'$ ($u'qq$, $u'u'q$, $u'\bar{q}$), which we denote as $h'$.
2) Form $B'=1$ baryons, composed of three $u'$.
3) Annihilate into SM particles via the formation of $u'\bar{u}'$.

The cross section of the rearrangement and the subsequent falling in the ground states is suppressed by the destruction of the excited states before falling. Taking this effect into account, the production cross section of the ground states is~\cite{DeLuca:2018mzn}
\begin{align}
\label{eq:h'Xsec}
\sigma \sim \frac{4\pi}{\Lambda_{\rm QCD}^2} \sqrt{\frac{\Lambda_{\rm QCD}}{m_{u'}}}.
\end{align}
The production cross section of $u'u'u'$ is of this order.
This is also effectively the annihilation cross section of $h'$ as  $u'\bar{u}'$ annihilate into SM particles.

\if0
In addition to the state $u'u'u'$, they are bounded into states including SM quarks, $u'u'q$, $u' qq$ and $u'\bar{q}$, which we collectively call as $h'$. The states $h'$ have fractional SM electromagnetic charges.
As the SM quarks are much more abundant than the mirror quarks, $u'$ first form bound stated with SM quarks with a large radius. They scatter with each other and produce excited bound states made only of $u'$ with large radii, $(u'\bar{u}')^*$ and $(u'u'u')^*$. The excited states can fall into the ground states with small radius. After taking into account the destruction of the excited states before falling into the ground states, the production cross section of the ground states is
%%
\begin{align}
\label{eq:h'Xsec}
\sigma \sim \frac{4\pi}{\Lambda_{\rm QCD}^2} \sqrt{\frac{\Lambda_{\rm QCD}}{m_{u'}}}.
\end{align}
%%
Once the ground states are formed, they are not destructed because of the large binding energy. The state $(u'\bar{u}')$ decays into a pair of gluons, and hence the production cross section of the state is effective the annihilation cross section of $h'$. The state $(u'u'u')$ is stable. The production cross section of it is also effectively the annihilation cross section of $h'$.
\fi

The abundances of $(u'u'u')$ and $h'$ is estimated as follows. If the cross section times the number density of $u'$ is larger than the Hubble expansion rate around the QCD phase transition, the abundance of $(u'u'u')$ is comparable to the initial abundance of $u'$. The abundance of $h'$ is given by the freeze-out abundance determined by the cross section in Eq.~(\ref{eq:h'Xsec}).
If the cross section is small, the abundance of $(u'u'u')$ is given by the freeze-in abundance, while that of $h'$ is close to the initial abundance of $u'$. The abundance of $(u'u'u')$ and $h'$ are given by
\begin{align}
Y_{u'u'u'} \simeq Y_{u'} \times
\begin{cases}
Y_{u'} / Y_{\rm crit} & Y_{u'} < Y_{\rm crit}\\
1 & Y_{u'} >Y_{\rm crit}
\end{cases}, \\
Y_{h'} \simeq Y_{u'} \times
\begin{cases}
1 & Y_{u'} < Y_{\rm crit}\\
 Y_{\rm crit} / Y_{u'}& Y_{u'} >Y_{\rm crit}
\end{cases},  \\
Y_{\rm crit} \equiv \left. \frac{H}{ \sigma v s} \right|_{T = \Lambda_{\rm QCD}} =10^{-16}   \frac{m_{u'}}{ 10^6 \, {\rm GeV}}.
\label{eq:Ycrit}
\end{align}
Ref.~\cite{Kawamura:2018kut} considers an alternative model where $U(1)_{EM}\times U(1)_{EM'}$ breaks to a single $U(1)_{EM}$ and additional scalar particles are introduced. Then $u'$ decays into a new particle and a SM quark. In their setup $e'$ is also unstable, and the additional scalar particles are dark matter candidates. We do not consider these non-minimal models in this paper.
%%%%%%%%%%%%%%%%%%
\subsection{The ICRR Limit on $u'$ Dark Matter}
\label{subsec:u'ICRR}
%%%%%%%%%%%%%%%%%%
The abundance of $h'$ is strongly constrained. Stringent constraints come from monopole searches of the 1980's, which are sensitive to ionization from fractionally charged $h'$. The bound from the ICRR experiment~\cite{Kajino:1984ug} is derived in~\cite{Dunsky:2018mqs} taking into account the acceleration by supernova remnants. For $m_{u'} = 10^{6-7}$ GeV the bound is $Y_{h'} < 10^{-25}$.%
\footnote{The bound is derived assuming that the charged particle does not feel strong interactions and may stop only from ionization losses in the atmosphere or Earth's crust. The ICRR experiment was situated above ground. Even with its strong interactions, we find $h'$ does not stop in the atmosphere nor the iron plates inside the ICCR detector for $m_{u'} = 10^{6-7} ~\GeV$.} 
This is much smaller than $Y_{\rm crit}$ of (\ref{eq:Ycrit}), so that the bound on the $u'$ abundance before the QCD phase transition is the same, $Y_{u'} < 10^{-25}$. The abundance of $u'u'u'$ is even smaller and almost all of DM is composed of $e'$.
Possible cosmological scenarios leading to the hierarchy of the abundances of $e'$ and $u'$ are discussed in Section~\ref{sec:nonthermal}.

%%%%%%%%%%%%%%%%%%
\subsection{Bulk Matter Constraints on $u'$ Dark Matter}
\label{subsec:u'bulkMatter}
%%%%%%%%%%%%%%%%%%

Additional constraints on $h'$ come from searches for fractionally charged particles in bulk matter, implemented via Millikan drop experiments or ferromagnetic levitometers \cite{Perl:2009zz}. While such experimental constraints are strong (no more than one $h'$ per $\sim 10^{21}$ nucleons) and mass-independent,
the results should be interpreted carefully, taking into account the distribution of $h'$ on Earth from billions of years of geologic churning, the potential contamination of the sample during the refinement process pre-experiment \cite{Lackner:1982dq}, and the uncertainty to what materials $h'$ may bind to due to the exotic chemistry of fractionally charged particles \cite{Lackner:1982dq,Lackner:1981jn}.
We (very) roughly estimate the relative number of $h'$ compared to nuclei in the crust as well as in meteorites and find that the flux constraints $f_{h'} \equiv \Omega_{h'}/\Omega_{DM} \lesssim 10^{-8}$ are already or marginally stringent enough to explain why $h'$ have gone undetected in such bulk matter experiments.

$h'$ which existed in the Earth before it solidified sank to the center of the Earth. Thus we consider $h'$ which has fallen onto the Earth after its solidification.
Supernova shocks partially evacuate $h'$ from the Milky Way disk so that the flux of both accelerated and unaccelerated $h'$ on Earth is approximately $\Phi \approx f_{h'} 10^5 ~\GeV/m_{h'} {~ \rm cm^{-2}s^{-1}}$ \cite{Dunsky:2018mqs}. The $h'$ impinging on the Earth with speed $v_{\rm{vir}}$ typically stop within a meter or so of crust, where geological effects become important. With typical geological denudation rates of order $v_{\rm churn} \sim 10^{-3} ~\rm{cm/yr}$ \cite{doi:10.1029/JZ069i016p03395}, a steady-state number density of $h'$ in the soil is reached with value
\begin{align}
	n_{h'} \sim \frac{\Phi}{v_{\rm churn}} \approx 	1  ~ {\rm cm^{-3}} \frac{f_{h'}}{10^{-8}}\frac{10^7 ~\GeV}{m_{h'}}.
	\label{eq:hdensitycrust}
\end{align}
The volume of each non-refined terrestrial sample tested for fractionally charged particles is $\lesssim 10^{-3} ~\rm{ cm^3}$ \cite{Perl:2009zz}, so that \eqref{eq:hdensitycrust} suggests fewer than one $h'$ resides in a given sample. It is thus highly plausible that $h'$ has escaped detection in such samples.

Bulk matter searches for fractionally charged particles have also been tested on meteorites which have the advantage of lacking the uncertainty associated with geological weathering. Moreover, iron meteorites are naturally ferromagnetic and hence can be minimally processed in principle before testing on ferromagnetic levitometers.

Meteorites are made of heavy elements which are synthesized in stars. As is argued in~\cite{DeLuca:2018mzn}, $h'$ are expected to sink toward the center of stars and annihilate, thereby reducing their abundance in meteorites. We thus consider the abundance of $h'$ in meteorites accumulated only during their exposure to cosmic rays, including $h'$. 

The distribution of $h'$ within the meteorite must be considered. For example, $h'$ with speed $v_{\rm{vir}}$ and charge $q e \approx 1$ impinging on the meteorite stop after $\sim 10 ~{\rm cm}$ and are thus typically ablated when the meteorite enters the atmosphere \cite{BHANDARI1980213,1968Metic...4..113M}. 
\footnote{If $m_{h'} \gtrsim 5/q^2 \times 10^8 ~\GeV$, $h'$ can pass right through even the largest sampled meteorite, Hoba, and hence avoid all bulk matter meteorite constraints. Note $|q|$ may be as low as $1/3$.}
Fermi-accelerated $h'$, on the other hand, can penetrate deeper into the core and avoid ablation losses. 
The accelerated spectrum of $h'$ induces a depth dependent number density within the meteorite. For low momentum, the Fermi-accelerated differential spectrum of $h'$, $d(n v)/dp = \Phi / p$, \cite{Dunsky:2018mqs}, so that the number density of $h'$ a distance $X$ below the meteorite surface is approximately 
\begin{align}
	n_{h'}(X) \sim \frac{\Phi \, t}{2 X} \approx 10^4 {~\rm cm^{-3}} \frac{f_{h'}}{10^{-8}}\frac{10^7 ~\GeV}{m_{h'}} \frac{0.5 \,{\rm m}}{X}\frac{t_{\rm CR}}{2 \times 10^8 ~{\rm yr}},
 	\label{eq:meteoriteDensity}
\end{align}
where $t_{\rm CR}$ is the exposure time of the meteorite to cosmic rays before falling to Earth.
We set the surface depth equal to the typical atmospheric ablation for meteorites like the Hoba sample, approximately $0.5 \,{\rm m} $.

The ablation length as well as the exposure time can be inferred by measuring the abundance of isotopes and the tracks of cosmic rays in a meteorite \cite{BHANDARI1980213}. For example, the Hoba meteorite experienced $40$ cm of ablation and about $2\times 10^8$ years of exposure to cosmic rays \cite{1968Metic...4..113M}. Since $10^{-4} ~{\rm cm^{3}}$ by volume of Hoba has been tested with null results~\cite{Jones:1989cq}, there is a good chance that no $h'$ are detected for $f_{h'} = 10^{-8}$. Besides Hoba, only three other meteorites have been tested, totaling less than $10^{-3} ~\rm{cm^{-3}}$ by volume \cite{Perl:2009zz,Jones:1989cq,Kim:2007zzs}. The exposure time to cosmic rays for each of these meteorites is far less than Hoba~\cite{BHANDARI1980213,fisher1960cosmogenic}, and thus give weaker constraints.

\subsection{Long-range self interaction of $e'$}
Mirror electrons interact with other mirror electrons via a massless mirror photon. Even though mirror electrons experience a long-range force, their mass is too heavy to appreciably self-scatter and disrupt the dark matter halo profile~\cite{Agrawal:2016quu} nor the spectrum of the cosmic microwave background.

%%%%%%%%%%%%%%%%%%
\subsection{The XENON1T Limit on $e'$ Dark Matter}
\label{subsec:e'}
%%%%%%%%%%%%%%%%%%

Mirror electrons also interact with SM particles via kinetic mixing and can produce an observable signal. The cross section of the scattering between $e'$ and a nucleus, of mass $m_N$ and atomic number $Z$, with relative velocity $v_{\rm rel}$ is given by
\begin{align}
\frac{d \sigma} {dq} = \frac{8\pi \alpha^2 Z^2 \epsilon^2}{v_{\rm rel}^2 q^3} |F(q)|^2,
\end{align}
where $q$ is the momentum transfer and $F(q)$ is the nuclear form factor. 
The number of expected events in a direct detection experiment with an energy threshold $E_{\rm th}$, a total target mass $M_{\rm tar}$, an exposure time $T$, and atomic weight $A$ is 
\begin{align}
N_{\rm event} = 1.6 \times  \left( \frac{\epsilon}{ 10^{-8}} \right)^2 \frac{10^7 \, {\rm GeV}}{m_{e'}}  \left( \frac{Z}{54} \right)^2 \left( \frac{131}{A} \right)^2 \frac{10{\rm \, keV}}{E_{\rm th}} \frac{f(E_{\rm th})}{0.3} \frac{M_{\rm tar} T}{{\rm ton}\times {\rm year}},
\end{align}
where we assume a local DM density of $0.3$ GeV/cm$^3$, as well as a velocity distribution of
\begin{align}
d v f(v) = dv  \frac{4}{\sqrt{\pi}} \frac{v^2}{v_0^3} \, {\rm exp}(- v^2/v_0^2),~~v_0= 220 \, {\rm km}/{\rm s}.
\end{align}
Here $f(E_{\rm th})$ takes into account the suppression of the scattering by the form factor,
\begin{align}
f(E_{\rm th}) = \left[ \int_{q_{\rm th}}^{q_{\rm max}} d q |F(q)|^2 q^{-3} \right] / \left[  \int_{q_{\rm th}}^{q_{\rm max}} d q q^{-3} \right], \nonumber \\
q_{\rm th} = \sqrt{2 m_N E_{\rm th}},~ q_{\rm max}= 2 m_N v_{\rm rel}.
\end{align}

XENON1T searches for a recoil between DM and Xenon with a threshold energy around 10 keV~\cite{Aprile:2018dbl}.
The bound obtained there can be interpreted  as an upper bound of 16 on the expected number of the events.
Assuming the Helm form factor~\cite{Helm:1956zz,Lewin:1995rx}, we find $f(E_{\rm th})\simeq 0.3$, so that the bound becomes
\begin{align}
m_{e'} > 1 \times 10^6 \, {\rm GeV} \left( \frac{\epsilon}{ 10^{-8}} \right)^2.
\end{align}
This result is translated to a bound in the $(v',\epsilon)$ plane in Fig.~\ref{fig:epsilon}.
Together with the prediction for $\epsilon$, this requires that the mirror electroweak scale is above $(3 \times 10^{11} -10^{12})$ GeV, for a UV cutoff ranging from $v'$ to $M_{Pl}$. The LZ experiment~\cite{Mount:2017qzi} is expected to provide about 10 times better sensitivity and probe $v'$ values an order of magnitude larger.  An experiment whose sensitivity is saturated by the neutrino-floor will have about 100 times better sensitivity~\cite{Billard:2013qya} and probe $v'$ values two orders of magnitude larger.  Note that larger values of $v'$ are expected to yield larger values of $\theta$ via the dimension 6 operator of (\ref{eq:HHGG}), as shown by vertical lines in Fig. \ref{fig:epsilon}, greatly enhancing the importance of the next 1-2 orders of magnitude of sensitivity in nuclear recoil experiments.

%%%%%%%%%%%%%%%%%%
\subsection{Correlations between $m_t, \alpha_s(m_Z)$ and the Direct Detection Rate}
\label{subsec:pred}
%%%%%%%%%%%%%%%%%%
The direct detection rate is a function of $v'$, which is determined by SM parameters. Future experiments will hone in on $v'$ and the direct detection rate as measurements of the top quark mass, strong coupling constant, and Higgs mass improve. The uncertainty on $v'$ comes dominantly from those of the top quark mass and the strong coupling constant.
We provide a fitting formula for $v'$ around $(m_t,\alpha_s(m_Z)) = (172.5~{\rm GeV}, 0.1192)$, 
\begin{align}
{\rm log}_{10}\frac{v'}{\rm GeV} \simeq 12.3 + 0.2 \left[ -   \frac{m_t -172.5~{\rm GeV}}{0.1~{\rm GeV}}  +  \frac{\alpha_s(m_Z)-0.1192}{0.0003}  +  \frac{m_h -125.18~{\rm GeV}}{0.18~{\rm GeV}}  \right] .
\label{eq:v'uncert}
\end{align} 
The uncertainty from the Higgs mass is sub-dominant, as seen in Fig. (\ref{fig:alphaSvsMtopPlot}).

\begin{figure}[tb]
\centering
\includegraphics[width=0.7\textwidth]{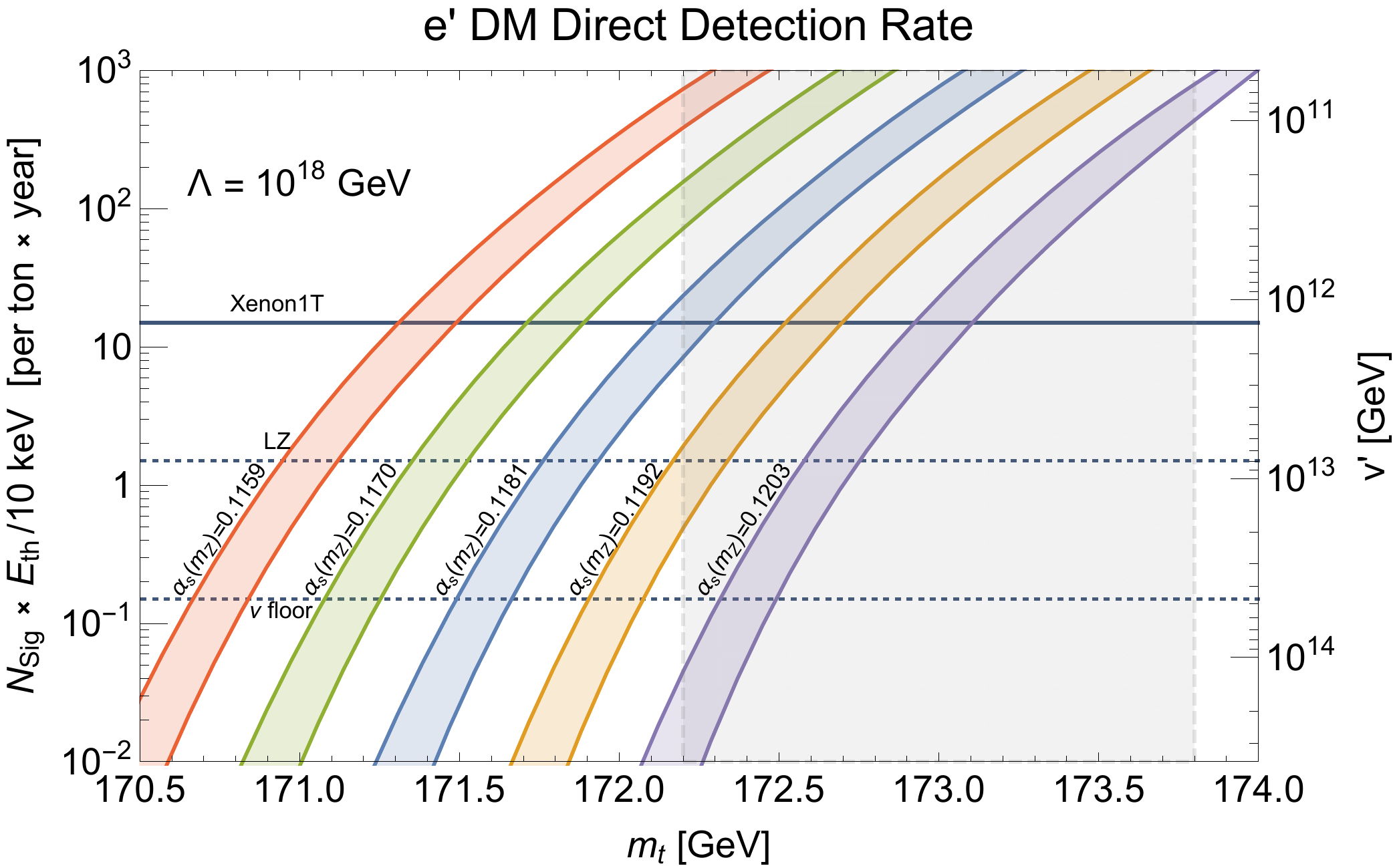}
\includegraphics[width=0.7\textwidth]{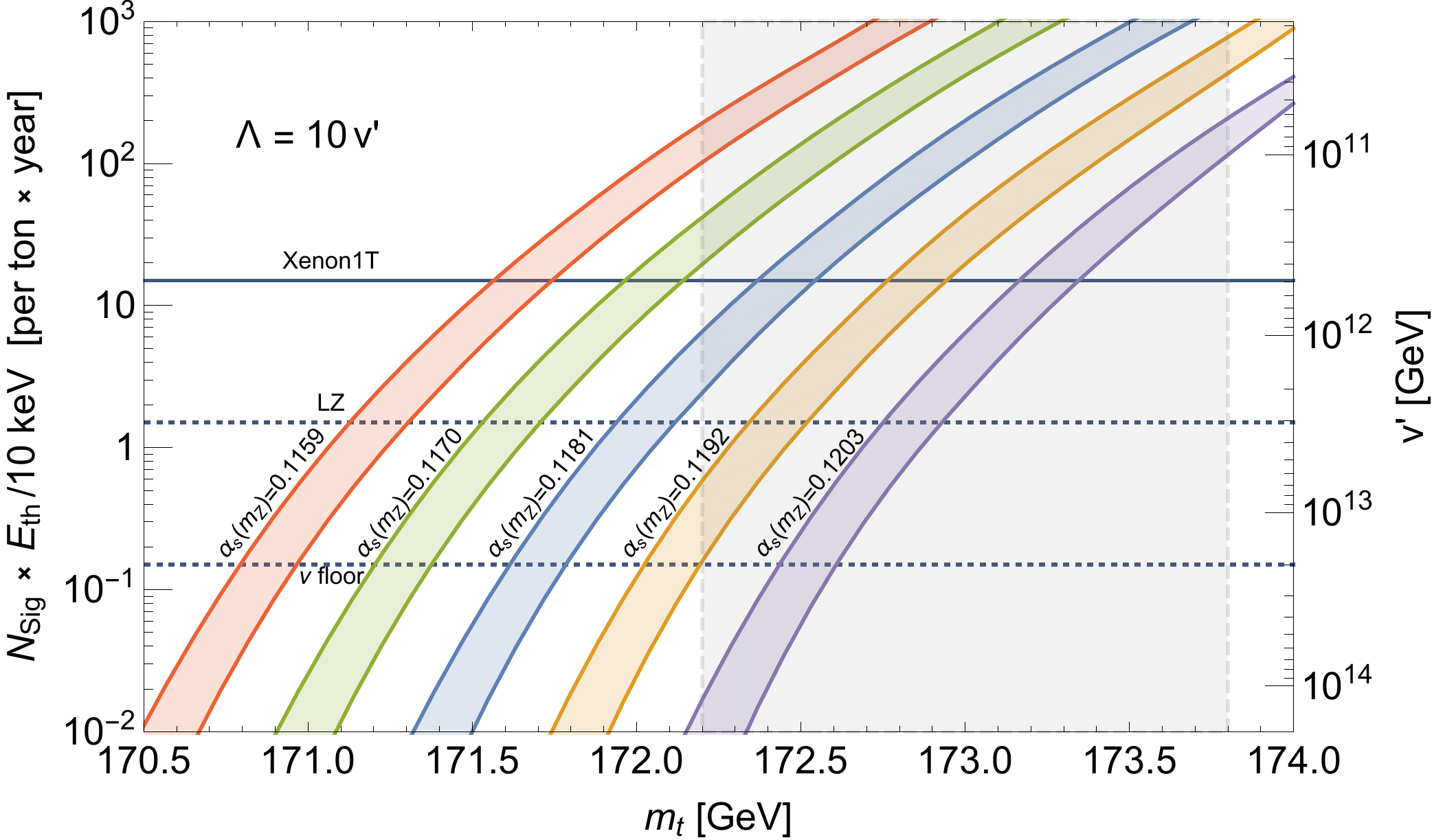}
\caption{The prediction for the $e'$ DM direct detection rate as a function of $m_t$. The thickness of the signal rate contours is due to the uncertainty in the Higgs boson mass. The gray shaded rectangle shows the current experimental value of $m_t$ to $2\sigma$.}
\label{fig:ddPlot}
\end{figure}

In Fig.~\ref{fig:ddPlot}, we show the prediction for the expected number of events, in experiments with Xenon targets, as a function of the top quark mass for a given strong coupling constant. We take a UV boundary condition for the kinetic mixing parameter of $\epsilon(\Lambda) = 0 $ with $\Lambda = 10^{18} ~\GeV$ ($10 \,v')$ in the upper (lower) panel, as shown by the green (blue) curves in Fig. \ref{fig:epsilon}. For a given set of the SM parameters, the difference in signal rates between these two cutoffs is only a factor of about 3 - 6.  The width of the bands correspond to the uncertainty from the Higgs mass. The horizontal solid line shows the bound from XENON 1T, while dashed lines show the sensitivity of future experiments. The constraint from XENON 1T already requires $m_t<173.1 \;(173.4)$ GeV.
 
The strong coupling constant can be measured with an accuracy of $0.1$\% by improving lattice computation as well as the conversion of the coupling at the lattice scale to that of higher energy scales~\cite{Lepage:2014fla}.   Further measurements at the $Z$-pole at lepton colliders can achieve similar accuracy~\cite{Gomez-Ceballos:2013zzn}. The uncertainty in the prediction of the event rate from the last term of (\ref{eq:v'uncert}) is then very small compared with that from the cutoff $\Lambda$.  The top quark mass can be measured with an accuracy of $0.2$ GeV at high-luminosity running at the LHC~\cite{CMS:2013wfa}, below which the uncertainty is saturated by the theoretical ambiguity associated with the definition of the pole mass and its conversion to $\overline{\rm MS}$~\cite{Bigi:1994em,Beneke:1994sw,Beneke:1998ui}. The Higgs mass can be determined with an accuracy of few 10 MeV at high luminosity running of the LHC~\cite{Cepeda:2019klc}.
At this stage the direct detection rate is predicted within a factor of about 6, where the uncertainty from the top quark mass dominates. Further improvement is possible by determining the $\overline{\rm MS}$ top quark mass directly by the measurement of the top quark production cross section which is free from the ambiguity. Lepton colliders can determine the top quark mass with an accuracy of few 10 MeV~\cite{Seidel:2013sqa,Horiguchi:2013wra,Kiyo:2015ooa,Beneke:2015kwa}, allowing for the prediction of the direct detection rate within few ten percents. With this accuracy, uncertainties from the local DM density, the velocity dispersion~\cite{Catena:2011kv,Bovy:2012tw}, the cutoff $\Lambda$, and the theoretical uncertainty in the determination of $v'$ become important.

%%

%%%%%%%%%%%%%%%%%%%%%%%%%%%%%%%%%%
\section{Cosmological Production of $e'$ Dark Matter}
\label{sec:nonthermal}
%%%%%%%%%%%%%%%%%%%%%%%%%%%%%%%%%
In this section we describe how the relic DM abundance is set in the early universe.
We assume that the matter-antimatter asymmetry of the mirror sector is negligible and discuss the abundance of symmetric components. This is the case, for example, if baryogenesis in the mirror sector is kinematically prevented because of the large mass scale of the mirror sector.

As we have seen in the previous section, the abundance of $e'$ must be much larger than that of $u'$. We first show that thermal production mechanisms do not work. 
The hierarchy of the abundances can be achieved by non-thermal production from the decay of the inflaton, or generically from a particle that dominates the energy density of the universe. This particle can have additional CP violating decay channels kinematically open to the SM but not the heavier, mirror sector, allowing a matter-antimatter asymmetry to develop solely in the SM.

%%%%%%%%%%%%%%%%%%
\subsection{Freeze-Out and Dilution}
\label{subsec:FO}
%%%%%%%%%%%%%%%%%%

For a large enough reheat temperature ($T_{RH} \gtrsim m_{u'}$), both the SM and mirror sectors are in thermal equilibrium. As a result, the abundance of $e'$ is set by thermal freeze-out and is given by $\Omega_{e'} \approx \Omega_{DM} (v'/10^8 \GeV)^2$
\footnote{This neglects the $e'$ produced from beta decays of heavier mirror fermions during $e'$ freeze-out, which exacerbates the overproduction problem.}
.
To produce the observed DM abundance, $v'$ is so low that it is already ruled out by kinetic mixing ($v' \lesssim 10^{12} ~\GeV$), besides predicting an unrealistically large top quark mass.

One way to increase $v'$ while maintaining $\Omega_{e'} = \Omega_{DM}$ is to dilute the frozen-out $e'$ by entropy produced from the decays of a massive particle that subsequently dominates the energy density of the universe. However, this fails because $e'$ and $u'$ have comparable freeze-out abundances and dilution changes their abundances by the same amount, preventing any hierarchy between $e'$ and $u'$ abundances from developing.
% \kh{I think we can consider generic entropy production and say that $u'$ is too abundant.}

%%%%%%%%%%%%%%%%%%
\subsection{Freeze-In}
\label{subsec:FI}
%%%%%%%%%%%%%%%%%%

Another potential thermal mechanism for producing $e'$ DM is through freeze-in from the SM plasma via electromagnetic interactions and kinetic-mixing, with $\epsilon \sim 10^{-8}$. 
Taking the reheat temperature after inflation below the $e'$ mass, $T_{RH} \ll m_{e'}$ leads to a freeze-in abundance with an exponential Boltzmann suppression, $\sim \exp(-2 m_{e'}/{T_{RH})}$.  The hope is that when this is chosen to give the observed DM abundance in $e'$, the heavier $u'$ will be even more Boltzmann suppressed so that its relic abundance is sufficiently small. However, $e'$ has the observed DM abundance if $T_{RH} \approx m_{e'}/10$ and, at this value of  $T_{RH}$, the freeze-in abundance of $u'$ is larger than for $e'$: the closeness of $m_{e'}$ and $m_{u'}$ means that the additional Boltzmann suppression of $u'$ production is more than compensated by the much stronger coupling of $u'$ to the SM via gluons. For the reasons discussed in Sec \ref{subsec:u'DM}, $u'$ must be highly sub-dominant relative to $e'$, hence the freeze-in origin for DM fails.

%%%%%%%%%%%%%%%%%%
\subsection{Non-Thermal Production from Decays of $\phi$}
\label{subsec:phidecay}
%%%%%%%%%%%%%%%%%%
 
We have seen that $u'$ is overproduced by many orders of magnitude in both freeze-out and freeze-in production of $e'$ DM. Nevertheless, non-thermal production of $e'$ DM from the decay of an inflaton $\phi$, (or any field which dominates the energy density of the universe), can produce $e'$ DM with a sufficiently small and innocuous abundance of $u'$ ($\Omega_{u'}/\Omega_{DM} \lesssim 10^{-8}$) if certain constraints on the inflaton reheat temperature and the $e'$ and $u'$ branching ratios are imposed.%
\footnote{It is also conceivable to produce $e'$ from a field whose energy density is subdominant when it decays.}
These general constraints are as follows:

First, the reheat temperature must be sufficiently low so that the  thermally produced freeze-in abundance of $u'$ from the SM bath is  $\lesssim 10^{-8} ~\Omega_{DM}$, implying
\footnote{Here we assume that the maximum temperature of the universe, $T_{\rm max}$, is $T_{\rm RH}$. If the decay is perturbative and the decay rate is constant in time, the maximum temperature is generically greater than $T_{\rm RH}$~\cite{Kolb:1990vq,Harigaya:2013vwa}. In this case the upper bound on $T_{\rm RH}$ is stronger. See~\cite{Harigaya:2014waa,Harigaya:2019tzu} for the estimation of DM abundance produced between $T_{\rm RH}$ and $T_{\rm max}$.}
\begin{align}
	T_{RH} \lesssim \frac{m_{u'}}{40 - \frac{1}{2}\ln (\frac{\Omega_{u'}/\Omega_{DM}}{10^{-9}})}.
%	T_{RH} \lesssim \frac{m_{u'}}{40}.
	\label{eq:tRHconstraint}
\end{align}
Next, decays of the inflaton must directly produce the observed DM abundance, requiring a branching ratio into $e'$ of
\begin{align}
B_{e'} \simeq\frac{{\rm eV}}{T_{RH}} \frac{m_{\phi}}{m_{e'}}
\label{eq:eBranching}.
\end{align}
Last, the inflaton branching ratio into $u'$ must be sufficiently small that $\Omega_{u'}/\Omega_{DM} \lesssim 10^{-8}$, implying
\begin{align}
	B_{u'} \lesssim 10^{-8} ~ B_{e'}.
	\label{eq:uBranching}
\end{align}
This small branching fraction requires $m_\phi$ to be in a narrow range, as it is challenging to obtain $B_{u'} \ll B_{e'}$ except by a kinematic suppression.

This seems to require a coincidence among the mass scales, which may be understood by an anthropic argument. Let us consider a landscape of vacua, scanning over the scale $v'$ while fixing other parameters of the theory.
Suppose that the structure of the theory is such that $u'$ is abundantly produced where kinematically allowed so that matter-radiation equality occurs much earlier than in our universe. A few examples are provided below. There are two possible obstacles for the formation of a habitable environment in such a DM-rich universe~\cite{Tegmark:2005dy}. First, the collapse of halos occurs much earlier, and hence galaxies are much denser than in our universe. A planet then has more frequent close encounters with stars, disturbing the habitable orbit around its own star. Second, the mass fraction of baryons is much smaller than ours. The baryons inside a disk are no longer self-gravitating and are stable against further collapse to form stars. Both obstacles require that the DM abundance should not exceed $O(10-100)$ times the DM abundance in our universe, so that universes with copious $u'$ production do not contain observers.%
\footnote{Note that we fix the magnitude of the primordial cosmic perturbation as well as the baryon density. The first and the second obstacles are avoided by decreasing the cosmic perturbation and increasing the baryon density, respectively.}
On the other hand, universes with $e'$ production kinematically forbidden have no DM. Almost no galaxies are formed before domination by dark energy, after which structure formation is prevented.

The requirements on $T_{RH}$, $B_{e'}$, and $B_{u'}$ described above can be satisfied, for example, in a model where the inflaton directly couples to quarks and gluons but not to leptons. To satisfy (\ref{eq:uBranching}), the upper bound on the inflaton mass is $m_\phi < 2 m_{u'}$. $e'$ DM is produced through decays $\phi \rightarrow \bar{e}' e' \gamma'$ via an off-shelf loop of mirror quarks and a virtual $\gamma'$. The inflaton coupling is determined so decays to quarks and gluons give $T_{RH}$ appropriately small to satisfy \eqref{eq:eBranching} and ensure that the freeze-in abundance of $e'$ is negligible.

Another model, which we will explore in detail in the future, can incorporate baryogenesis. The inflaton directly couples to heavy right-handed neutrinos $N, N'$, that are integrated out to yield dimension 5 operators of (\ref{eq:high_dimnu}), leading to masses for the neutrinos $\nu$ and $\nu'$. The inflaton decays to $\nu'$ via the mixing between the right-handed neutrinos and $\nu'$. The beta decay of $\nu'$ into $e'\bar{e}'$ and a lighter $\nu'$, which is suppressed by the large mirror electroweak scale, produces $e'$ DM with a small branching ratio. The decay into $u'$ is forbidden by imposing $m_{\nu'} < m_{u'} + m_{d'} + m_{e'}$. The anthropic argument is applicable if the beta decay of $\nu'$ into $e'\bar{e}'$ and a lighter $\nu'$ involves a small mirror MNS angle.
$\nu'$ also decays into the Standard Model left-handed leptons and the Higgs, and leptogenesis~\cite{Fukugita:1986hr} occurs non-thermally~\cite{Lazarides:1991wu,Asaka:1999yd}. 
A SM matter-antimatter asymmetry is generated via the interference between the tree and one-loop decay diagram of $\nu'$ via the operator $l' lH$, akin to the decay of sterile neutrinos via the operator $N lH$. However, there is no mirror matter-antimatter asymmetry since the large mass scale of the mirror sector prevents an analogous reaction.
 
%%%%%%%%%%%%%%%%%%
\section{Conclusions and Discussions}
%%%%%%%%%%%%%%%%%%
\label{sec:conclusions}
The Standard Model is remarkable: it correctly describes a wide wealth of data, while giving a highly incomplete understanding of particle physics.  At its inception, there was an immediate realization that one must seek a deeper theory beyond.  A particularly elegant idea is to unify the three gauge forces \cite{Georgi:1974sy,Georgi:1974yf}, despite their manifest differences.  Furthermore, if there is a desert above the weak scale, $v$, the unification of couplings at a very large energy scale $M_G$ leads to a prediction for the proton decay rate, $\Gamma_p$
\begin{align}
	\{\alpha_i\} \;\; \rightarrow \;\; \frac{M_G}{v}, \hspace{1in} \Gamma_p \propto \frac{1}{M_G^4}.
	\label{eq:GUTs}
\end{align}

In the intervening decades, despite a succession of ever more powerful experimental tests, the Standard Model, with three generations, neutrino masses and a single Higgs doublet, has shown ever wider applicability.  We are motivated to pursue an alternative completion far in the UV because the observed value of the Higgs mass implies that the SM possesses another scale, $\mu_c$, where the Higgs quartic coupling vanishes 
\begin{align}
	\{\alpha_i, m_t, m_h \} \;\; \rightarrow \;\; \frac{\mu_c}{v}, %\hspace{1in} \Gamma_p \propto \frac{1}{M_G^4}.
	\label{eq:newscale}
\end{align}
and we take the view that this is the next symmetry breaking scale of nature.  Which deeper symmetries of nature should be introduced and broken at $\mu_c$? Motivated by the strong CP problem we introduce a Higgs Parity that includes spacetime parity but does not replicate QCD, and motivated by DM we introduce mirror electroweak gauge symmetry.

We have constructed the minimal theory with gauge group $SU(3) \times SU(2) \times U(1) \times SU(2)' \times U(1)'$ with Higgs Parity exchanging the two electroweak groups and the corresponding two Higgs doublets, $H$ and $H'$.  The new symmetry breaking is accomplished by $\vev{H'}=v'$, which is a mirror version of the SM electroweak breaking $SU(2)' \times U(1)' \rightarrow U(1)_{EM'}$, with $v'\simeq \mu_c$. Remarkably, this theory has {\it the same number of parameters as the SM} while solving the strong CP problem and providing a DM candidate, the mirror electron $e'$.  In addition, a very small kinetic mixing parameter results from a 4-loop gauge calculation and provides the interaction between $e'$ and ordinary matter that allows a prediction of the event rate $N_{\rm event}$ at nuclear recoil direct detection experiments   
 \begin{align}
	\{\alpha_i, m_t, m_h \} \;\; \rightarrow \;\; \frac{v'}{v}, \hspace{1in} N_{\rm event} \propto \frac{1}{v'}.
	\label{eq:Higgsparityunif}
\end{align}
 
We comment on the comparison between grand unification (\ref{eq:GUTs}) and our UV completion of the SM (\ref{eq:Higgsparityunif}).  Both have a compelling signal with a rate suppressed by the high symmetry breaking scale, $M_G$ for proton decay, and $v'$ for DM direct detection.  A succession of experiments was necessary to reduce the uncertainties on $\{\alpha_i\}$ so that $M_G$, and hence the proton decay rate, could be precisely predicted.  This was made difficult because $\Gamma_p$ depends on the 4th power of $M_G$.  Although the minimal theory is excluded, unified threshold corrections allow more complicated models.  Similarly, in the theory of this paper further experiments are now needed  to better measure $\{\alpha_s, m_t, m_h \}$ to pin down $v'$ and hence the direct detection rate.  Here one is greatly aided by two features: $N_{\rm event}$ falls only linearly with $v'$, and there is a second observable, the neutron electric dipole moment, that grows as $v'^2$. Figure \ref{fig:ddPlot} shows that, no matter how the values of $\{\alpha_s, m_t, m_h \}$ evolve as uncertainties are reduced, the entire parameter space of the theory will be tested.  As in grand unification, adding particles in the desert could destroy the prediction; however, extra particles added at the scale $v'$ do not easily affect our prediction.  There is an uncertainty coming from the UV completion scale for the calculation of the kinetic mixing parameter, but this is a logarithmic effect that leads at most to an uncertainty of 2.5 around the central prediction.   Unlike minimal grand unification, our theory implies that the gauge structure gets more complicated before any ultimate simple unification.

\let\oldaddcontentsline\addcontentsline% Store \addcontentsline
\renewcommand{\addcontentsline}[3]{}% Make \addcontentsline a no-op

\section*{Acknowledgement}
We thank Nima Arkani-Hamed, Takemichi Okui and Satoshi Shirai for useful discussion.
This work was supported in part by the Director, Office of Science, Office of High Energy and Nuclear Physics, of the US Department of Energy under Contracts DE-AC02-05CH11231 and DE-SC0009988 (KH), as well as by the National Science Foundation under grants PHY-1316783 and PHY-1521446. 

\let\addcontentsline\oldaddcontentsline% Restore \addcontentsline

\appendix

\bibliography{epDM}

\let\addcontentsline\oldaddcontentsline% Restore \addcontentsline
  
\end{document}